\documentclass[lettersize,journal]{IEEEtran}
\usepackage{amsmath,amsfonts}
\usepackage{algorithmic}
\usepackage{array}
\usepackage{textcomp}
\usepackage{stfloats}
\usepackage{url}
\usepackage{upgreek}
\usepackage{verbatim}
\usepackage{svg}
\usepackage{graphicx}
\usepackage{tikz}
\usepackage[caption=false,font=footnotesize]{subfig} % in der Präambel

\usepackage{acro}

\DeclareAcronym{dpd}{
  short = DPD ,
  long  = digital predistortion
}

\def\BibTeX{{\rm B\kern-.05em{\sc i\kern-.025em b}\kern-.08em
    T\kern-.1667em\lower.7ex\hbox{E}\kern-.125emX}}
\usepackage{balance}
\begin{document}
\title{Digital Predistortion of Power Amplifiers {for} Quantum Computing}
\author{Marvin Jaeger ,~\IEEEmembership{Graduate Student Member,~IEEE}, Bartosz Tegowski, Georg Frederik Riemschneider,\\ and Alexander Koelpin,~\IEEEmembership{Fellow,~IEEE}

\thanks{ 
This work is part of the Hamburg Quantum Computing (HQC) project, where the required technologies for the implementation of a quantum computer are developed. The project is co-financed by ERDF and Fonds of the Hamburg Ministry of Science, Research, Equalities and Districts (BWFGB). (\textit{Corresponding author: Marvin Jaeger}.)\\
The authors are with the Institute of High-Frequency Technology, Hamburg University of Technology, Hamburg, Germany
(e-mail: marvin.jaeger@tuhh.de).\\
This work has been submitted to the IEEE for possible publication. Copyright may be transferred without notice, after which this version may no longer be accessible.}
}

%\markboth{IEEE MICROWAVE AND WIRELESS TECHNOLOGY LETTERS}%
%{How to Use the IEEEtran \LaTeX \ Templates}

\maketitle

\bstctlcite{IEEEexample:BSTcontrol}
\begin{abstract}
Power amplifiers (PA) are essential for microwave-controlled trapped-ion and semiconductor spin based quantum computers (QC). They adjust the power level of the control signal and therefore the processing time of the QC. Their {nonlinearities and memory effects} degrade the signal quality and, thus, the fidelity of qubit gate operations. Driving the PA with a significant input power back-off reduces nonlinear effects but is neither power-efficient nor cost-effective. To overcome this limitation, this letter augments the conventional signal generation system applied in QCs by digital predistortion (DPD) {to linearize the radio frequency (RF) channel}. {Numerical analysis of the qubit behavior based on measured representative control signals} indicates that DPD improves its fidelity.

\end{abstract}


\section{Introduction}
\IEEEPARstart{Q}{uantum} computers (QC) are predicted to be superior to classical computers in solving probability-based problems. Unlike classical computers, which use the distinct states ``0'' and ``1'', QCs use the superposition of the quantum states $\lvert 0 \rangle$ and $\lvert 1 \rangle$, which forms the qubit state. There exist different physical implementations of qubits, such as superconducting qubits, trapped ions, and semiconductor spin qubits, whose quantum states can be manipulated by using microwave control signals \cite{MicrowaveQC,MicrowaveQC2}. Therefore, the quality of the microwave control signals directly affects the qubit state fidelity of quantum operation \cite{qubitFidelity}. The signal quality depends on the signal generation system, where the power amplifier (PA) ensures a sufficient output power level. The PA, however, is a nonlinear device with memory elements, which leads to amplitude errors, time delay, phase shifts, and spurious generation, thereby decreasing the fidelity. In quantum computing, a common method to compensate for those nonideal effects is to increase the input power back-off (IBO), which requires PAs with higher output powers to maintain a specified power level to drive the qubit gates. This increases costs and decreases power efficiency.

\begin{figure}[]
	\centering
	\footnotesize
	\def\svgwidth{.62\textwidth}
	%\hspace*{2cm}
	\includesvg{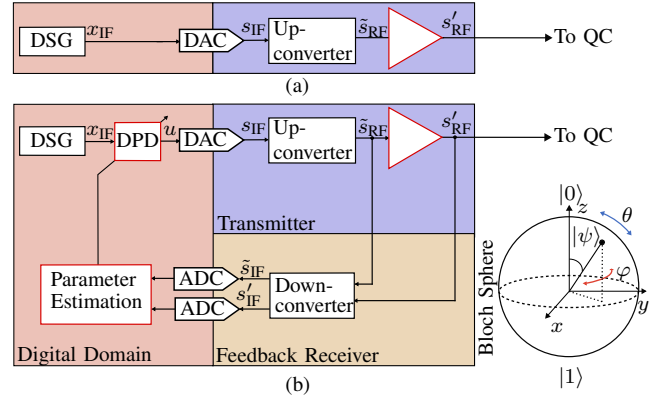}
	\caption{{Conceptual transmitter architecture of (a) a classical quantum control system and (b) a quantum control system extended by DPD and a feedback loop for PA characterization. The Bloch sphere visualizes the qubit state $\lvert \psi \rangle$.}}
	
	\label{fig:QCS}
	\vspace{-5mm}
\end{figure}

Instead of increasing the IBO, the {nonlinearities and memory effects} can be compensated. {Early linearization relied on analog feedforward and feedback techniques. However, with today’s high bandwidths and adaptation requirements, digital predistortion (DPD) has become established in communication engineering \cite{PALinJournal}}, {while analog predistortion \cite{APD} represents a viable alternative that can provide wideband linearization at high carrier frequencies with inherently low latency, albeit with reduced flexibility for reconfiguration}. In DPD, the inverse of the nonlinear transfer function is modeled in the digital domain and applied to the input signal to linearize the overall system. Commonly used regression-based methods are the {generalized memory polynomial (GMP)} \cite{DPD_GMP} and its parameter-reduced form, i.e., the memory polynomial (MP)~\cite{DPD_MP}. Different neural-network-based DPD approaches have also been reported in the literature \cite{DPD_ANN,DPD_LSTM,WuRL}.
{All linearization methods involve a tradeoff between computational complexity and the achievable performance of the power amplifier (PA), in terms of power efficiency and linearity. }

{In quantum-control systems, signals are either precomputed and replayed using an arbitrary waveform generator \cite{Krantz2019QuantumEngineerGuide} or real-time calculated based on given waveform parameters \cite{Qick}.} {Computational complexity and processing latency are relevant for live waveform generation but do not affect relative gate timing.} {Achieving accurate wideband DPD requires sufficient bandwidth in the digital-to-analog converters (DACs), analog-to-digital converters (ADCs), and frequency-conversion stages.} {Existing quantum-control signal generators, such as the QICK platform \cite{Qick}, already consists of high-performance components, enabling the integration of DPD with only limited additional system adaptation.}

This letter studies the application of DPD to the control signal generation dedicated for microwave-controlled QCs. {For this, the classical transmitter architecture \cite{Qick}, \cite{GEMIC25}, shown in Fig.~\ref{fig:QCS}(a), is augmented by a feedback receiver and DPD functionality as indicated in Fig.~\ref{fig:QCS}(b).} By compensating for the {nonlinearities and memory effects} of the PA, the control signal accuracy can be improved. Numerical simulations of qubit state evolution based on measured control signals indicate an improvement in fidelity without compromising the output power. {In quantum-control systems, linear equalization, pulse shaping, and error-compensating control pulses are widely used and actively researched \cite{Krantz2019QuantumEngineerGuide}}. {However, to the best of the authors’ knowledge, the explicit application of DPD to quantum control signal generation has not yet been reported.}

\section{Qubit Gate Operation} \label{sec2}
In a QC, a single qubit state can be expressed as~\cite[eq.~(1)]{MicrowaveQC2}
\begin{equation}
   \lvert \psi \rangle = \alpha \lvert 0 \rangle + \beta \lvert 1 \rangle, \label{eq:qubit}
\end{equation}
where $\alpha$ and $\beta$ are complex-valued probability amplitudes normalized by $|\alpha|^2 + |\beta|^2 = 1$. It can be represented in the Bloch sphere with the corresponding angles $\theta$ and $\varphi$, as depicted in Fig.~\ref{fig:QCS}(b). A defined rotation in $\theta$ and $\varphi$ forms a qubit gate operation, { represented by the state evolution matrix $U$, which changes the initial qubit state $\lvert \psi \rangle$ to the final state~\cite[eq.~(7.2)]{lapierre2021introduction}
\begin{equation}
	\lvert \psi' \rangle = U\cdot\lvert \psi \rangle. \label{eq:psi}
\end{equation}

The qubit gate can be numerically approximated by \cite[eq.~(1)]{qubitFidelity}
\begin{equation}
	U \approx \prod_{n=N}^{0} \exp\left(-jH(n \Delta t)\Delta t\right), \label{eq:U}
\end{equation}
where $H$ denotes the Hamiltonian, $n$ the discrete time index, $\Delta t$ the time step, and $\exp(\cdot)$ the matrix exponential. Equation~\eqref{eq:U} approximates the time-varying Hamiltonian by a product of time-independent components.
For single-qubit gate operation, the Hamiltonian in the rotating frame under the rotary wave approximation is defined as \cite[eq.~(91)]{Krantz2019QuantumEngineerGuide}
\begin{equation}
	H(n \Delta t) = -\frac{\Omega_R}{2} \left(
	\Re\{ s_\text{BB}(n \Delta t) \} \sigma_x
	+ \Im\{ s_\text{BB}(n \Delta t) \} \sigma_y
	\right),\label{eq:H}
\end{equation}
where $\sigma_x$ and $\sigma_y$ are the Pauli matrices and $s_\text{BB}(t)$ is the equivalent baseband representation of the microwave qubit excitation signal. A sufficiently small time step $\Delta t$ is required to ensure an accurate numerical approximation of the Hamiltonian.

The qubit state fidelity \cite[eq. (1)]{stateFidelity}
\begin{equation}
	F
	= \left| \langle \psi_\text{ideal} \mid \psi \rangle \right|^2 \label{eq:F}
\end{equation}
evaluates how close the qubit state $\lvert \psi \rangle$ is to the ideal gate $\lvert \psi_\text{ideal} \rangle$. 

}

\section{Quantum Control Signal Generation}
For qubit implementations based on trapped-ions or semiconductor spin qubit types considered in this work, the qubit state can be adjusted by a microwave signal whose frequency corresponds to the qubit excitation frequency $f_e$. The signal amplitude is proportional to the Rabi frequency $\Omega_R$, which is the angular speed of rotation along $\theta$ and mainly impacts the QC processing speed. The phase of the signal controls the angle $\varphi$. Simultaneous processing of multiple qubits can be realized by spatial or frequency multiplexing.  For example, this is exploited e.g. by the magnetic gradient induced coupling (MAGIC) principle \cite{MAGIC}.

To excite $N$ frequency multiplexed qubits for $M_n$ consecutive gates, the multitone signal
\begin{equation}
	s_{\text{RF}}(t) = \sum_{n=0}^{N-1}s_{\text{RF},n}(t)   \label{eq:excitation}
\end{equation}
is applied, where each tone
\begin{equation}
	s_{\text{RF},n}(t) = \sum_{m=0}^{M_n-1} \Re\left\{s_{\text{BB}{m,n}}(t) \cdot e^{j2\pi f_{e,n} t}\right\} \label{eq:gates}
\end{equation}
with
%\underbrace{A_{m,n}\cdot p_{m,n}(t-\tau_{m,n})}_{s_{\text{BB}{m,n}}(t)}
\begin{equation}
	s_{\text{BB}{m,n}}(t) = A_{m,n}\cdot p_{m,n}(t-\tau_{m,n}) e^{j \phi_{m,n}}
\end{equation}
individually controls its associated qubit, which exhibits the excitation frequency $f_{e,n}$. $A_{m,n}$ denotes the amplitude of the signal, $\phi_{m,n}$ is the signal phase, and $p_{m,n}(t)$ is the pulse form applied for each gate at $\tau_{m,n}$, which can be summarized to the baseband signal $s_{\text{BB}{m,n}}(t)$. For a rectangular pulse of duration $T$, the $\theta$-rotation in the Bloch sphere amounts to $\theta$\,=\,$\Omega_R\cdot T \propto A_{m,n} \cdot T$. 
 
\begin{figure}[]
	\centering
	\hspace{2.3mm}
	\vspace{-1mm}
	\def\svgwidth{.5\textwidth}
	\includesvg{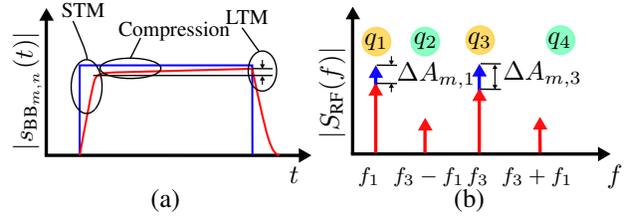}
	\caption{{(a) Ideal (blue) and distorted (red) magnitude of the equivalent baseband output signal. (b) Spectrum illustrating desired qubit excitations (yellow) and idle qubits (green).}}
	\label{fig:qubit_spect}
%	\vspace{-3mm}
\end{figure}
{Fig.~\ref{fig:QCS}(a) shows a typical generation system \cite{Qick,GEMIC25}, which intend to output the signal \eqref{eq:excitation}.} It is generated as the equivalent baseband (BB) signal $s_{BB_{m,n}}$ and upconverted to the intermediate frequency~(IF) band signal $x_\text{IF}$ using a {digital signal generator (DSG)} in the digital domain. A DAC converts the digital signal into the analog domain.
The transmitter transforms this IF signal $s_\text{IF}$ into the required radio frequency~(RF) signal $s_\text{RF}$ by upconversion and amplification as in~\cite{GEMIC25}. The RF band is determined by the excitation frequency of the qubits.

Compared to the upconverter, the PA introduces the dominant signal distortion in the transmitter chain, which degrades the signal quality.  The characteristics of the PA can be described as
\begin{equation}
	s_{\text{RF}}'(t) = G_\text{lin} \cdot h\{\Tilde{s}_\text{RF}(t)\}, \label{eq:pa_tf}
\end{equation}
where $G_\text{lin}$ is the linear gain and $h\{\cdot\}$ is the dynamic transfer function, which accounts for {nonlinearities and memory effects}. {The nonlinear effects are due to saturation and lead to amplitude compression (AM/AM) and amplitude-dependent phase distortion (AM/PM).} The memory effects can be separated into long-term memory (LTM) and short-term memory~(STM)~\cite{PA_Book}. LTM is caused by thermal effects, supply modulation, trapping, and device aging, while STM is due to parasitic circuit components.

Fig.~\ref{fig:qubit_spect}(a) depicts the influence of STM and LTM on the equivalent BB signal in time domain. Due to compression, the output amplitude is reduced. STM causes increased switch-on and switch-off times of the output signal which cause a time delay of the pulse. LTM causes an increase of the amplitude during the excitation due to heating up the PA. Fig.~\ref{fig:qubit_spect}(b) shows a qualitative output spectrum of a dual tone signal used to intentionally excite the qubits $q_1$ and $q_3$ (yellow). Compression reduces the amplitude by causing an error $\Delta A_{m,n}$. In addition, the nonlinear effects introduce intermodulation products, which, in this example, unintentionally excite $q_2$ that is supposed to idle (green). Furthermore, the intermodulation product at $f_3+f_1$ leads to partial excitation of qubit $q_4$ due to spectral proximity.

%%%----------------------------------------------------------------------------------------------------------------------------------------------------------------------------------------------------------
These effects can be mitigated by applying DPD in the digital domain, where the intermediate-frequency signal $s_\text{IF}$ is predistorted to obtain the output signal $u$, as shown in Fig.~\ref{fig:QCS}(b). To model the inverse of the PA transfer function $h^{-1}\{\cdot\}$ required for DPD, the classical signal generation system is extended by a measurement of the in- and the output signal of the PA. For this, a feedback receiver is introduced, which downconverts the PA's input and output signals $\Tilde{s}_\text{RF}$ and ${s}_\text{RF}$ to the IF signals $\Tilde{s}_\text{IF}$ and $s'_\text{IF}$, respectively, where they are digitized by an ADC. In the digital domain, a parameter estimation algorithm estimates $h^{-1}\{\cdot\}$ and adjusts the DPD, by using, e.g., a least squares estimator or optimization techniques. {The achievable DPD performance is limited by the accuracy of the feedback path}.

%------------------------------------------------------------------------------

\section{Experimental Results}\label{sec3}%%%%%%%%%%%%%%%%%%%%%%%%%%%%%%%%%%%%%%%
To evaluate the effectiveness of DPD in the context of quantum computing, the qubit gate fidelity is numerically simulated based on measured qubit control signals without and with DPD applied.

\subsection{Experimental Setup}
%system

The measurements of representative control signals are performed on the RF WebLab platform \cite{rfweblab2025}, which consists of a vector signal transceiver (National Instruments PXIe-5646R) and a gallium nitride PA (Cree CGH40006-TB). Even though the system is not dedicated for the control of a QC, it implements the signal generation system depicted in Fig.~\ref{fig:QCS}(b). Thus, it is representative to study the impairments of the microwave signal quality, the consequences of which can be discussed in terms of QC operation by means of numerical qubit simulations according to \eqref{eq:psi}--\eqref{eq:F}. The system has a measured linear gain of 25.48\,dB and supports a 200-MHz bandwidth at 2.14-GHz center frequency. The received signal is synchronized by adding a maximum length sequence to the signal, upsampling by a factor of {10}, applying a circular cross correlation, {and downsampling by the same factor. }

%signalaufbau
{
To demonstrate the effectiveness of DPD, signals for frequency-multiplexed qubits are used, which strongly suffer from nonlinear effects due to their high dynamic range.
As an example, $N$\,=\,4 BB signals are considered, where each one is devoted to one simulated qubit at 2.16, 2.17, 2.18, and 2.19\,GHz. The signal is separated in sequences, which consists of four pulses: two $\pi$-pulses and two idle-pulses. Each pulse has an equal duration of $T$\,=\,1.6\,$\upmu$s. The order of the pulses is randomly selected and individual for each qubit excitation. The $\pi$-pulse is randomly chosen between $\pm X_\pi$, and $\pm Y_\pi$, where $X$ and $Y$ denote a rotation around the $x$- and $y$-axis in the Bloch sphere, respectively. Each sequence ideally implements an identity operation, and 20 subsequent sequences are applied. To suppress image frequencies, a single-sideband (SSB) modulation is applied during the signal generation. The transmitted signal $s_\text{IF}$ has an {average output power} of $P_n\,\text{=}\,-22.5$\,dBm per tone, which is assigned to a maximum Rabi frequency of $\Omega_{R,\text{max}}$.}

%DPD
{As an example among different DPD algorithms, the MP is used, which is defined as \cite{DPD_MP}
\begin{equation}
	u
	= \sum_{k=1}^{K} \sum_{l=0}^{L-1}
	a_{k,l}\, x_\text{IF}(n-l)\, \lvert x_\text{IF}(n-l) \rvert^{k-1},
\end{equation}
where $a_{k,l}$ are the coefficients weighting the nonlinear order $k$ and memory depth $l$, estimated by minimizing the mean square error over 50 sequences. In the experiments, $K=5$ and $L=6$ are used. }

\begin{figure}[]
	\centering
	
	\footnotesize
	\def\svgwidth{.5\textwidth}
	\includesvg{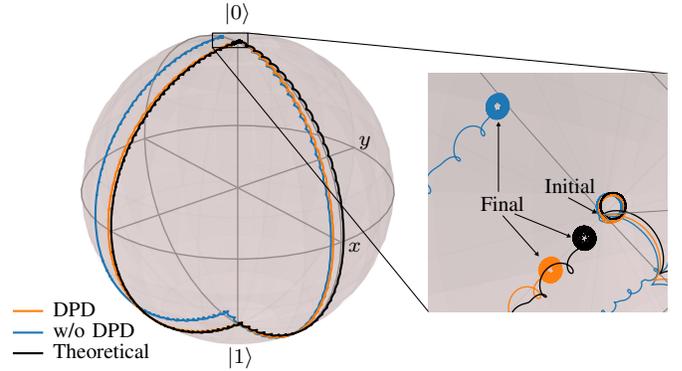}
	\caption{{Calculated trajectory of an exemplary sequence (idle,~$Y_\pi$,~$-X_\pi$,~idle) within the Bloch sphere, calculated for qubit~1 based on measured control signals.}}
	\label{fig:bloch}
%	\vspace{-5mm}
\end{figure}
\begin{figure}[]
	\centering
	\vspace{-4mm}
	\footnotesize
	\def\svgwidth{.5\textwidth}
	\includesvg{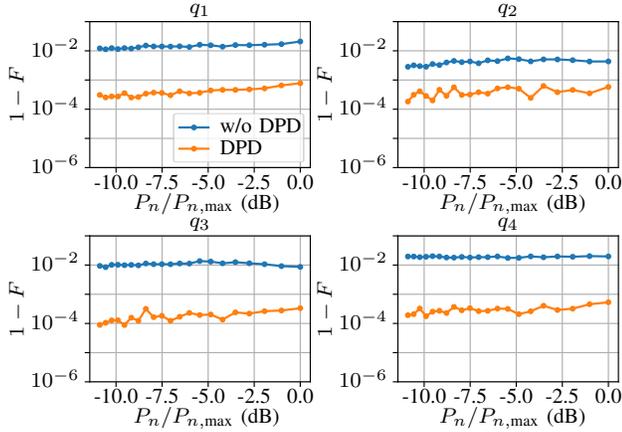}
	\caption{{Mean infidelity versus normalized average output power of the different qubits, swept over a set of output powers.}}
	\label{fig:sequence}
	\vspace{-5mm}
\end{figure}

{
%-----------------------------------------------------------------------------------------------
\subsection{State-Fidelity Analysis of an Exemplary Sequence} 
Fig.~\ref{fig:bloch} exemplarily illustrates the simulated state vector trajectory of qubit~1 of the 15th  sequence. It consists of an idle period, a $Y_\pi$-pulse, a $-X_\pi$-pulse, and a subsequent idle period. The trajectory is calculated according to~\eqref{eq:U}. For this, the Hamiltonian~\eqref{eq:H} is evaluated for the sequence, based on the qubit's equivalent BB signal $s_{BB}$, which is determined by digitally downconverting the measured signal $s_\text{IF}'$ to the BB domain of the respective qubit in accordance with its frequency offset and is normalized to the range $\pm$1. The first driving pulse within the first sequence is used to determine the static signal phase to initialize the qubit state. The four qubits are considered as four independent single-qubit systems with individual excitation frequencies but subject to the same control signal. 

%Thus, the numerical simulation of the state evolution accounts for amplitude and phase errors of the control signal as well as intermodulation products near the excitation frequencies.

According to Fig.~3, a qubit phase shift in $\varphi$ due to LTM is observed with and without DPD.
The oscillations on the state trajectory are caused by the interference from the excitation tones of the other qubits, which is also presented in the theoretical trajectory, which directly derives from $x_\text{IF}$.
Comparing the final state with the desired state, using~\eqref{eq:F}, leads to a fidelity of $F$\,=\,99.4\,\%. With and without DPD, over-rotation errors accumulate over subsequent pulses, whereas DPD mitigates this effect, resulting in a higher fidelity of $F$\,=\,99.81\,\% for this exemplary sequence. 

\vspace{-3mm}
\subsection{Analysis Over Output Power}
To generalize the observation, the fidelity is evaluated over a set of output powers. For a decreased average output power $P_n$, the pulse duration is increased by the same factor to maintain $\pi$-pulse generation. The same sequences are executed for each power level and repeated ten times to suppress measurement noise, both without and with DPD applied. The DPD coefficients are recalculated for every output power level.

Fig.~\ref{fig:sequence} presents the mean infidelity $1-F$ for every qubit depending on the relative power level. Even for an ideal PA, the qubit gates would be nonideal due to the interference caused by the excitation signals of the neighboring qubits. Thus, to keep the focus on the fidelity related to {nonlinearities and memory effects} of the PA, the fidelity with and without DPD is calculated with respect to the theoretical state that results from the signal $x_\text{IF}$ instead of the ideal state $\lvert \psi_\text{ideal} \rangle$. Fig.~\ref{fig:sequence} indicates that DPD reduces the infidelity overall. Note that the infidelity without DPD nearly remains constant over the sweep because the linear gain is remeasured each power level step for a fair comparison.
}

\section{Conclusion}\label{sec4}
This letter outlines the mitigation of quantum gate errors of microwave-controlled QCs due to nonlinearity and memory effects of the PA in their control signal generators. By compensating for nonlinear and memory effects using DPD, the accuracy of the control signal can be improved. For this, a feedback loop is introduced to characterize the PA transfer function. {Numerical qubit behavior simulations based on measured control signals of a representative signal generation system indicate an improved quantum state fidelity thanks to DPD.} Consequently, the application of DPD enables increased processing speeds of QCs without compromising quantum state fidelity, thereby increasing power efficiency.

%\vfill\break

\bibliographystyle{IEEEtran}

\bstctlcite{IEEEexample:BSTcontrol}
\bibliography{MyBib,BSTcontrol.bib}

\end{document}